\documentclass[aps,prx,superscriptaddress,twocolumn,floatfix]{revtex4-2}
\usepackage[utf8]{inputenc}
\usepackage{amsmath, bm}
\usepackage{cancel}
\usepackage{soul}
\usepackage{amsfonts}
\usepackage[normalem]{ulem}
\usepackage{mathtools}
\usepackage{physics}
\usepackage{dsfont}
\usepackage{verbatim}

\usepackage{color}
\usepackage[dvipsnames]{xcolor}
\usepackage{textgreek}
\usepackage{amssymb}
\usepackage{eucal}
\usepackage{tcolorbox}
\usepackage{wasysym}
\usepackage{float}
\usepackage{subfigure}
\usepackage{amsthm, amssymb,bbm}
\usepackage{booktabs}
\usepackage{array}
\usepackage[colorlinks=true,citecolor=Cerulean,linkcolor=RubineRed,urlcolor=Cerulean]{hyperref}

\usepackage{algorithm}
\usepackage{algpseudocode}
\usepackage{float}     

\newcommand{\bseq}{\begin{subequations}}
\newcommand{\eseq}{\end{subequations}}

\begin{document}

\title{Uncovering and Circumventing Noise in Quantum Algorithms via Metastability} 

\author{Antonio Sannia}
\email{sannia@ifisc.uib-csic.es}
\affiliation{Quantum and Condensed Matter Physics Group (T-4), Theoretical Division, Los Alamos National Laboratory, Los Alamos, New Mexico 87545, USA}
\affiliation{Institute for Cross-Disciplinary Physics and Complex Systems (IFISC) UIB-CSIC, Campus Universitat Illes Balears, 07122,
Palma de Mallorca, Spain}
\affiliation{USRA Research Institute for Advanced Computer Science (RIACS), USA} 
\author{Pratik Sathe}
\affiliation{Quantum and Condensed Matter Physics Group (T-4), Theoretical Division, Los Alamos National Laboratory, Los Alamos, New Mexico 87545, USA}
\affiliation{Information Science \& Technology Institute, Los Alamos National Laboratory, Los Alamos, NM 87545, USA}
\author{Luis Pedro Garc\'ia-Pintos}
\affiliation{Quantum and Condensed Matter Physics Group (T-4), Theoretical Division, Los Alamos National Laboratory, Los Alamos, New Mexico 87545, USA}

\begin{abstract}
    The presence of noise is the primary challenge in realizing fault-tolerant quantum computers. In this work, we introduce and experimentally validate a novel strategy to circumvent noise by exploiting the phenomenon of metastability, where a dynamical system exhibits a separation of time scales in its evolution. We demonstrate that if quantum hardware noise exhibits metastability, both digital and analog algorithms can be designed in a noise-aware fashion to achieve intrinsic resilience. We develop a general theoretical framework and introduce an efficiently computable noise vulnerability metric that avoids the need for full classical simulation of the quantum algorithm. We show that the noise vulnerability index bounds errors in noisy implementations, with smaller values indicating greater fidelity between the achieved and target quantum states. We illustrate the use of our framework with applications to variational quantum algorithms and analog adiabatic state preparation. Crucially, we provide experimental evidence supporting the presence of metastable noise in gate-model quantum processors and quantum annealing devices. Thus, we establish that the noise properties in near-term quantum hardware can directly inform practical implementation strategies, enabling the preparation of final noisy states that more closely approximate the ideal ones.
\end{abstract}

\maketitle

\section{Introduction}

Noise and decoherence remain the primary obstacles for experimentally demonstrating quantum advantages over classical methods~\cite{Preskill2018quantumcomputingin,StilckFrana2021}. Despite significant advances in hardware design, current error rates remain too high to enable large-scale fault-tolerant computation. Consequently, developing strategies to mitigate or even exploit noise has become a central goal in quantum information science~\cite{Noise_RevModPhy}. In this work, we introduce a novel approach to characterizing and alleviating the effects of noise in quantum algorithms. Our key insight is that noise can exhibit a structured behavior that can be harnessed to protect quantum computations. In particular, we focus on metastability~\cite{Brinkman2022}, a phenomenon in which a dynamical system evolves on well-separated time scales.

Metastability has been extensively studied in classical contexts, including statistical physics~\cite{Langer1969}, chemical systems~\cite{Chem}, and neuroscience~\cite{Tognoli2014}. Interestingly, it also arises in quantum systems~\cite{Metastability_QS} and has been experimentally observed in platforms such as neutral atoms~\cite{Wu2022, Chen2022, Darbha2024}, ion traps~\cite{Allcock2021}, and superconducting qubits~\cite{qubit_metastable}, with promising theoretical proposals in quantum algorithms, including quantum associative memory~\cite{Associative_Memory, LabayMora2025} and quantum error correction~\cite{Botzung2025}.

Based on this concept, we show that if quantum hardware noise induces metastability, both digital and analog algorithms can be designed in a noise-aware fashion, achieving intrinsic resilience without requiring redundant encoding. This approach differs from conventional strategies, such as quantum error correction~\cite{lidar2013quantum,Terhal2015} and decoherence-free subspaces~\cite{Lidar1998,Lidar2014}, which rely on adding extra qubits to encode information, introducing the experimental challenge of implementing non-transverse operations. While recent works have explored similar directions to tackling decoherence~\cite{funcke2024robustness, LuisPedroGarcaPintos2025, Berberich2025, berberich2025robustness, zeng2025fundamentalcostsnoiserobustquantum}, a major limitation remains: existing methods lack an efficient way to compute a noise resilience metric. Typically, they require full classical simulation of the quantum algorithm, thereby preventing simultaneous evaluation of noise resilience and the attainment of quantum advantage.

To address this shortcoming, we introduce a noise vulnerability measure, $\mathcal{V}$, that, under standard assumptions on the noise model, can be efficiently computed for a wide class of algorithms without requiring complete knowledge of the algorithm's output. Importantly, we show that computing $\mathcal{V}$ enables the derivation of a worst-case fidelity bound between the ideal algorithmic output and its noisy counterpart. Moreover, in addition to presenting a general theoretical framework, we illustrate our theory with applications to well-known algorithms such as variational quantum algorithms~\cite{Peruzzo2014, Cerezo2021} and adiabatic state preparation~\cite{asp, Albash2018}. Importantly, our work extends beyond theoretical proposals--- we present experimental evidence supporting the presence of metastable noise for IBM's superconducting devices and D-Wave's quantum annealers~\cite{JohnsonQuantum2011}. These results suggest that metastability can be directly leveraged to enhance algorithmic performance on near-term quantum hardware.

\vspace{-0.5cm}

\section{Metastability}

In the context of dynamical systems, metastability refers to the emergence of intermediate, long-lived states resulting from the interplay of multiple dynamical timescales~\cite{Brinkman2022}. This phenomenon plays a central role in the dynamics of open quantum systems, which we consider in the following~\cite{Metastability_QS}. Under the Markovian approximation, the general evolution of a density matrix $\rho$ describing such a system is governed by the Gorini–Kossakowski–Lindblad–Sudarshan (GKLS) master equation~\cite{breuer2002theory, gorini1976completely, lindblad1976generators}:
\begin{align}
\label{eq:Lindblad}
\frac{d\rho}{dt} &= \mathcal{L}[\rho] \equiv -i[H, \rho]
+ \sum_{i} \gamma_i \Bigl(
    L_i \rho L_i^\dagger - \frac{1}{2} \{ L_i^\dagger L_i, \rho \}
\Bigr) \\
    &= (\mathcal{U} + \mathcal{D})[\rho] \,, \nonumber
\end{align}
where $H$ is the system Hamiltonian (generating the unitary superoperator $\mathcal{U}$), $\{L_i\}$ are the Lindblad (jump) operators modeling the coupling to the environment, and $\{\gamma_i\}$ are the associated decay rates. The dissipative contribution is collected in the superoperator $\mathcal{D}$.

Metastability in this setting is intimately connected to the spectral properties of the non-Hermitian Liouvillian superoperator $\mathcal{L}$. For a system of $n$ qubits, and neglecting possible exceptional points~\cite{exc_points}, the Liouvillian can be diagonalized in a biorthogonal basis of left and right eigenmatrices, $\{ \ell_j \}$ and $\{ r_j \}$, such that
\begin{equation}
\mathcal{L}[r_j] = \lambda_j r_j, \quad
\mathcal{L}^\dagger[\ell_j] = \lambda_j^* \ell_j, \quad
\Tr\{\ell_j^\dagger r_k\} = \delta_{jk}
\end{equation}
where all eigenvalues $\{\lambda_j\}$ satisfy $\Re(\lambda_j) \leq 0$ due to the contractivity of quantum channels.

For definiteness, suppose $\mathcal{L}$ admits a unique stationary state $\rho_{\mathrm{ss}}$ with $\mathcal{L}[\rho_{\mathrm{ss}}] = 0$. Then, any initial state $\rho(0)$ evolves as:
\begin{equation}
\label{Eq:evol}
\rho(t) = \rho_{\mathrm{ss}} + \sum_{j\geq1} e^{\lambda_j t}\, \Tr\{\ell_j \rho(0)\}\, r_j \,.
\end{equation}
All non-stationary contributions decay with characteristic timescales $\tau_j = 1/|\Re(\lambda_j)|$ and oscillate with frequencies $\omega_j = |\Im(\lambda_j)|$, ultimately relaxing the system to $\rho_{\mathrm{ss}}$. When there is a clear separation between these timescales, metastability arises~\cite{Metastability_QS}. For instance, if $\tau_{i} \ll t \ll \tau_{j}$ for some indexes $i$ and $j$, the fast modes have decayed while slower modes appear confined to a metastable manifold spanned by those right eigenvectors $r_m$ whose eigenvalues satisfy $|\Re(\lambda_{m})| \leq 1/\tau_{j}$. If there is additionally a separation of timescales in the coherent dynamics such that $ t' \ll 1/\omega_{i'},  1/\omega_{j'}$ for some indexes $i'$ and $j'$ belonging to the metastable manifold, then the system appears stationary in this intermediate timescale. Metastability also implies that different initial conditions $\rho(0)$ approach the stationary state $\rho_{\mathrm{ss}}$ at rates determined by the projection onto the various decay modes.

Assuming that the noise affecting the performance of quantum algorithms exhibits metastability, manifested through clearly separated and observable timescales, we show how quantum algorithms can be adapted to exploit this structure. By leveraging the hierarchy of noise timescales, it is possible to operate in regimes where rapid components are eliminated, thereby improving robustness and overall performance.

\section{Noisy quantum circuits}
\subsection{Noise resilience index}
A generic digital quantum algorithm can be described as a sequence of $L$ layers of unitary operations $\{U_k\}$, typically implemented by a quantum circuit, applied to an initial state $\rho_{\mathrm{in}}$. An ideal, noiseless implementation ends with a final state $\rho_f^{\mathrm{ideal}}$, that is given by
\begin{equation}
\label{Eq:Ideal}
\rho_f^{\mathrm{ideal}} = \Bigl( \prod_{k=1}^L U_k \Bigr)\, \rho_{\mathrm{in}}\, \Bigl( \prod_{k=1}^L U_k \Bigr)^\dagger.
\end{equation}
In practice, quantum circuits are affected by noise, which can be modeled by Markovian quantum channels represented as $e^{\mathcal{L}_k}$, where the evolution time is absorbed into the Liouvillian superoperator $\mathcal{L}_k$ acting after each unitary. The resulting noisy final state $\rho_f^{\mathrm{noisy}}$ is thus
\begin{equation}
\label{Eq:Noise}
\rho_f^\text{noisy} = \Lambda_L \circ \dotsc \Lambda_{k} \dotsc  \circ \Lambda_1 [\rho_\text{in}], 
\end{equation}
where $\Lambda_k [\mathcal \rho] = e^{\mathcal L_k} [U_k \rho U_k^\dagger]$.

To analyze the effects of noise, we expand each $\mathcal{L}_k$ in its eigenbasis: let $\{ r_{i_k}^k \}$ denote the right eigenmatrices of $\mathcal{L}_k$ and $\{ \lambda_{i_k}^k \}$ their eigenvalues. Then the final noisy state of Eq.~\eqref{Eq:Noise} can be written as
\begin{equation}
\label{Eq:RhoN}
\rho_f^{\mathrm{noisy}} = \sum_{i_1, \ldots, i_L}
\beta_{i_1,\dotsc, i_L} \,
e^{ \sum_{k=1}^L \lambda_{i_k}^k }
\, r_{i_L}^L,
\end{equation}
where the expansion coefficients
$\beta_{i_1,\dotsc,i_L}
=
\alpha_{i_1}^1 \alpha_{i_1,i_2}^2 \cdots \alpha_{i_{L-1},i_L}^L$
are defined recursively through the decompositions
$U_1 \rho_{\mathrm{in}} U_1^\dagger
=
\sum_{i_1} \alpha_{i_1}^1 r_{i_1}^1$
and
$U_k\, r_{i_{k-1}}^{k-1} U_k^\dagger
=
\sum_{i_k} \alpha_{i_{k-1},i_k}^k r_{i_k}^k$
for $k=2,\ldots,L$.

The noiseless outcome $\rho_f^{\mathrm{ideal}}$ is immediately recovered by formally setting all exponential factors $e^{ \sum_k \lambda_{i_k}^k } \to 1$ in Eq.~\eqref{Eq:RhoN}. Consequently, given a set of noise generators $\{\mathcal{L}_k\}$, optimizing the noise resilience of a quantum algorithm corresponds to minimizing the contributions associated with these exponential terms.

To assess an algorithm's vulnerability to noise, we introduce the index
\begin{align}\label{Eq:R}
    \mathcal{V}=\mathcal{V}(\{U_k\})  = \max_{\substack{i_1, \ldots, i_L \\ \beta_{i_1\dotsc i_L}\neq 0}} \abs{1- e^{\sum_{k=1}^L \lambda_{i_k}^k }}, 
\end{align}
where the maximum is evaluated over the non-zero terms in Eq.~\eqref{Eq:RhoN} and it is  upper bounded by an algorithm-independent index $\mathcal V_0$:
\begin{align}
    \mathcal{V} \leq \mathcal V_0 = \max_{i_1, \ldots, i_L} \abs{1- e^{\sum_{k=1}^L \lambda_{i_k}^k }}.
\end{align}

Focusing on this quantity highlights the most vulnerable component of the dynamics, irrespective of the amplitude of its associated eigenvector. This approach fundamentally differs from other metrics in the literature that quantify the impact of decoherent noise~\cite{funcke2024robustness, LuisPedroGarcaPintos2025, berberich2025robustness}, which, being based on fidelity calculations, require knowledge of the final state of the quantum algorithm. As shown below, the strength of our method in contrast lies in the fact that $\mathcal{V}$ can be efficiently computed for a broad class of algorithms without assuming the ability to simulate the total quantum evolution, thereby preserving the possibility of a genuine quantum advantage.

Importantly, as we will show, $\mathcal{V}$ admits a clear operational interpretation, as it directly determines a bound on the fidelity $F$ between the ideal and noisy final states. In particular, for the class of noise-informed quantum circuits considered here, the deviation from perfect fidelity is upper-bounded by $\mathcal{V}$. That is,
\begin{align}
F(\rho_f^{\text{ideal}}, \rho_f^{\text{noisy}})
\ge 1 - \mathcal{V} \geq 1 - \mathcal V_0.
\end{align}

We note that, although the full set of eigenvalues 
$\{\lambda^k_{i_k}\}_{i_k}$ at each layer is fixed by the noise model, 
the specific algorithm determines which sequences of noise eigenmodes contribute 
to the expansion in Eq.~\eqref{Eq:RhoN}. Consequently, the exponents 
that enter the maximization defining $\mathcal{V}$, and therefore 
$\mathcal{V}$ itself, can vary across different algorithms.
Moreover, distinct algorithms can share the same maximum value of $\mathcal{V}$; however, counting how many times this maximum appears in the final state decomposition of Eq.~\eqref{Eq:RhoN} provides a way to discriminate their noise vulnerability. 

Importantly, the possibility of tuning $\mathcal{V}$ across different quantum algorithms can be exploited only if there is a separation of time scales in the underlying noise channels, which corresponds to the presence of metastability. If these time scales were not well separated, the exponential term in Eq.~\eqref{Eq:R} would not differ significantly for distinct non-vanishing values of the $\beta_{i_1, \dots, i_L}$ coefficients.

Finally, as shown in the Supplementary Material, the definition of $\mathcal{V}$ can be generalized to the broad setting of analog algorithms, including quantum protocols with explicit dissipative dynamics~\cite{Diehl2008,Verstraete2009,Harrington2022,Sannia2024dissipationas,Sannia2024BP,Mi2024,maciejewski2024improving}. In this case, both the algorithm and the noise are generated by time-dependent Liouvillians, $\mathcal{L}_I(t)$ and $\mathcal{L}_N(t)$, respectively.

\begin{figure}[t!]
    \centering  
    \includegraphics[width=0.8\linewidth,keepaspectratio]{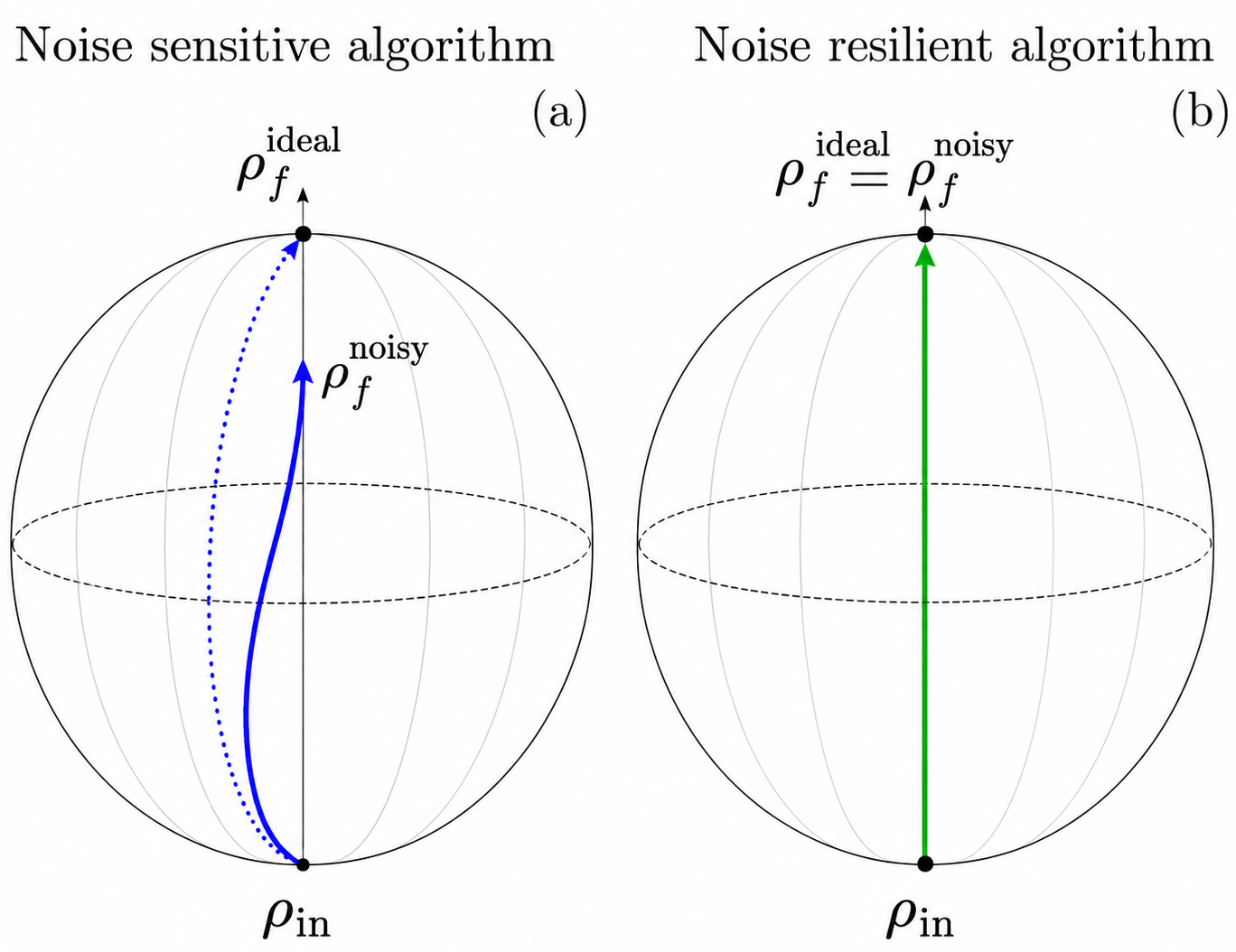}
    \caption{Representation of the single-qubit illustrative example on the Bloch sphere. (a) State preparation via rotation around the x-axis. The dotted line shows the ideal trajectory, while the solid line represents the noisy evolution, which deviates because of noise-induced decay. (b) State preparation through the amplitude damping channel. The states remain on the z-axis, fully mitigating the effects of noise.}
    \label{fig:Algs}
\end{figure}

\section{Single-qubit illustrative example}

We begin by illustrating our framework with a simple single-qubit example. Specifically, we consider the task of preparing the qubit in the state $\ket{0}$ starting from $\ket{1}$. We model the noise as a Pauli-diagonal Lindblad superoperator of the form
\begin{equation}\label{Eq:unb_Lindb}
\mathcal{D} = \gamma_x \mathcal{D}_x + \gamma_y \mathcal{D}_y + \gamma_z \mathcal{D}_z,
\end{equation}
where $\mathcal{D}_a$ denotes the dissipator associated with the jump operator $\sigma^a$ $(a \in {x,y,z})$. We focus on a biased noise model with $\gamma_z = \gamma$, while $\gamma_x = \gamma_y = 0$, i.e., pure decoherence along z. In this regime, states aligned along the $z$-axis of the Bloch sphere are unaffected by the noise.

As a first approach, we consider an analog algorithm implementing a bit-flip operation,
\begin{equation}
\ket{0} = U(T)\ket{1},
\end{equation}
where $U(T)$ is a unitary rotation around the $x$-axis of the Bloch sphere with evolution time $T = \pi$.

Considering a weak noise limit $\gamma \ll 1$, as shown in the Supplementary Material, the final noisy state takes the form
\begin{align}
\rho^N_{f,x}
&\simeq \tfrac12\big(\mathbb{I} + e^{-\gamma T}\sigma^z\big),
\end{align}
which yields the following fidelity with respect to the final ideal state
\begin{align}
F_x(\rho^N_{f,x}, \ket{0}\bra{0})
\simeq  \tfrac12\big(1 + e^{-\gamma T}\big)=1 - \frac{\mathcal{V}_x}{2}.
\end{align}
Here, $\mathcal{V}_x$ denotes the noise vulnerability index associated with this protocol.

As an alternative strategy, we consider a dissipative state-preparation scheme in which $\ket{0}\bra{0}$ is prepared through an amplitude-damping process, allowing the system to relax toward the desired target state. In this case, the evolution remains confined to the $z$-axis, and the noisy protocol coincides with the ideal one. Consequently, the related noise-vulnerability index is $\mathcal{V}_z = 0$ with a corresponding fidelity $F_z = 1$. In terms of our metastability framework, this example represents an extreme case in which the evolution generated by the most resilient algorithm produces states whose associated decay rates vanish identically, corresponding to an infinite decay timescale and thus to zero noise vulnerability.

\vspace{0.5cm}

\section{Computing the noise vulnerability index for quantum circuits}

In general, the full tomography of a quantum channel demands resources that grow exponentially with the number of qubits. In contrast, practical descriptions of the noise affecting real quantum circuits typically require only a polynomial amount of resources. In particular, Pauli-twirling techniques yield compact approximate 
Pauli-diagonal representations of the effective noise channels 
$e^{\mathcal{L}_k}$ appearing in Eq.~\eqref{Eq:Noise}~\cite{Emerson2007,Magesan2011,Cai2019,Erhard2019,Inter_RB}. 
Namely, each channel can be written as
\begin{equation}\label{Eq:Pauli_dec}
    e^{\mathcal{L}_k}[\rho]
    =
    \sum_{i=0}^{N_k} p_{k,i}
    \tilde{P}_{k,i}\rho \tilde{P}_{k,i},
\end{equation}

where $\{\tilde{P}_{k,i}\}$ is the set of Pauli strings describing the noise action at the $k$-th layer. In particular, $\tilde{P}_{k,0}=\mathbb{I}$ is the identity operator and, for $i\geq 1$, $\tilde{P}_{k,i}$ can be a generic $n$-qubits Pauli string. The coefficients $p_{k,i}\geq 0$ are the 
corresponding probabilities and satisfy $\sum_{i=0}^{N_k} p_{k,i}=1$. The number of retained non-identity Pauli strings typically scales polynomially with the number of qubits, $N_k=\mathcal{O}(\mathrm{poly}(n))$.

For this noise model, the whole set of Pauli strings form an eigenbasis 
of the corresponding Liouvillian superoperator. Using the notation 
introduced above, the associated right and 
left eigenvectors can then be chosen as
\begin{equation}
    r^k_{i_k}=P_{i_k},
    \qquad
    \ell^k_{i_k}=\frac{P_{i_k}}{2^n},
\end{equation}
so that $\mathrm{Tr}\{(\ell^k_{i_k})^\dagger r^k_{j_k}\}
=\delta_{i_k j_k}$. We denote the corresponding Liouvillian eigenvalues by 
$\lambda^k_{i_k}$, which can readily be computed from Eq.~\eqref{Eq:Pauli_dec}.

This decomposition enables the efficient computation of $\mathcal{V}$ for a 
broad class of quantum algorithms. For simplicity, we focus here on the case 
in which the eigenvalues $\lambda^k_{i_k}$ are real. Like in the case of the hardware efficient ansatzes studied below, we consider circuits where each unitary $U_k$ can be factorized into a product of a Clifford gate $U_{c,k}$ and a non-Clifford gate 
$U_{nc,k}$, such that $U_k=U_{nc,k} U_{c,k}$.

At each layer $k$, let $\mathcal{F}_k$ denote the set of populated Pauli 
strings associated with the fastest decaying noise modes, namely
\begin{equation}
    \mathcal{F}_k
    =
    \left\{
    P_i \,:\, 
    \lambda^k_i
    =
    \min_{j\in \mathcal{A}_k} \lambda^k_j =\lambda_f^k
    \right\},
\end{equation}
where $\mathcal{A}_k$ is the set of Pauli strings that generate the
algorithm output at layer $k$. Since the eigenvalues are real and non-positive, the 
minimum eigenvalue corresponds to the fastest decay.

The key assumption is that these fastest-decaying sectors can be efficiently 
tracked throughout the circuit. More explicitly, as in the following examples, we assume that the Clifford layers map the relevant 
fastest-decaying set at layer $k-1$ into the corresponding set at layer $k$,
\begin{equation}
    U_{c,k}\mathcal{F}_{k-1}U_{c,k}^\dagger
    \subseteq
    \mathcal{F}_k.
\end{equation}
Moreover, we also assume that the non-Clifford gates preserve
a nonzero overlap with it. In particular, we require
\begin{equation}
\Pi_{\mathcal{F}_k}
\!\left[
U_{nc,k} P_i U_{nc,k}^\dagger
\right]
\neq 0,
\qquad
\forall P_i \in \mathcal{F}_k ,
\tag{18}
\end{equation}
where \(\Pi_{\mathcal{F}_k}\) is the projector onto \(\mathcal{F}_k\). 

Under these conditions, the dominant exponential contribution in 
Eq.~\eqref{Eq:RhoN} can be identified without reconstructing the full Pauli 
decomposition of the state. Consequently, $\mathcal{V}$ can be computed by 
following only the noise modes that govern the largest decay, determined by the action of the Clifford operators, thereby avoiding 
a classical simulation of the full noisy quantum evolution.

Importantly, this estimation does not imply that the output of the 
quantum algorithm can be efficiently determined. In fact, the presence of non-Clifford operations renders, in general, the circuit classically hard to simulate.

Assuming that the fastest decay terms after the first layer are known, as for the hardware efficient ansatzes here studied, the procedure can be summarized as follows:
\begin{algorithm}[H]
\caption{Computation of the noise vulnerability index}
\label{alg:Noise_res}
\begin{algorithmic}[1]
    \Require $\rho_{\mathrm{in}}$ \Comment{Initial state}
    \Require $\{U_{c,k}\}$ \Comment{Set of Clifford gates}
    \Require $\{\tilde{P}_{k,i}\}$ \Comment{Noise Pauli strings}
    \Require $\{p_{k,i}\}$ \Comment{Associated Pauli probabilities}
    \Ensure $\mathcal{V}$ \Comment{Noise resilience index}
    \State Initialize: $\lambda_f \gets 0$
    \State Get the first layer fastest decay modes: 
    \[
        U_{1}\rho_{\mathrm{in}}U^{\dagger}_{1} \rightarrow \ \mathcal{F}_1
    \]
    $ \lambda_f \gets \lambda_f + \lambda^1_f$  
    \For{$k = 2$ \textbf{to} $L$} 
        \State Update the fastest decay modes under the Clifford action:
        \[
            \mathcal{F}_k \gets 
            U_{c,k} \, \mathcal{F}_{k-1} \, U_{c,k}^\dagger
        \]
        \State Match $\mathcal{F}_k$ with its corresponding eigenvalue $\lambda^{k}_{f}$
        \State Update:
        \[
            \lambda_f \gets \lambda_f + \lambda^{k}_{f}
        \]
    \EndFor
    \State \Return $\mathcal{V} =1-e^{\lambda_f}$
\end{algorithmic}
\end{algorithm}

From a computational-complexity perspective, the cost of the procedure is uniquely determined by the action of the Clifford unitaries, each of which can be performed in $O(n^2)$ time~\cite{Aaronson2004}. Therefore, for a circuit with $L$ layers, the total runtime scales as $O(Ln^2)$. We stress that the algorithm's computational complexity is independent of $N_k$. Consequently, even in a hypothetical scenario in which a Pauli twirling procedure does not scale efficiently, once the noise model has been characterized, the computation of the noise vulnerability remains unaffected.

The reason we are able to efficiently compute the noise vulnerability is that, for the class of circuits considered, we can efficiently access the fastest decaying modes at each layer. Consequently, this result readily generalizes beyond the setting of Eq.~\eqref{Eq:Pauli_dec}. For example, as we will show below, for noise models derived from measured qubit decay times $T_1$ and $T_2$, commonly used to describe noise in quantum circuits, the complexity of the procedure is not altered. Moreover, under similar hypothesis, in the Supplementary Material, we discuss how $\mathcal{V}$ can be upper bounded in the general case of analog algorithms.

\section{Fidelity bound for quantum circuits}

We now quantitatively relate the noise vulnerability index \(\mathcal V\) to the closeness between the noisy and ideal output states by deriving a lower bound on the fidelity. We first consider the class of noise-informed circuits, namely circuits designed to minimize the effect of noise at each layer. For such circuits, the final noisy state can be written as
\begin{equation}\label{Eq:Pauli_met_exp}
    \rho_f^{\mathrm{noisy}} =
    \frac{\mathbb I}{2^n}
    + e^{\sum_{k=1}^L\lambda_f^k} \sum_i \alpha_{i} P_{i}
    =
    \frac{\mathbb I}{2^n}
    + e^{\lambda_f} \sum_i \alpha_{i} P_{i},
\end{equation}
where \(\{P_i\}\) denotes the full set of Pauli strings, \(\{\alpha_i\}\) are the corresponding expansion coefficients, and
\(\lambda_f=\sum_{k=1}^L\lambda_f^k\). For noise-informed circuits, the decay rate selected at each layer satisfies
\begin{equation}
    \lambda_f^k = \max_{i_k}\lambda_{i_k}^k .
\end{equation}

The ideal final state is recovered in the noiseless limit \(\lambda_f \to 0\). Since \(\rho_f^{\mathrm{ideal}}\) is assumed to be pure, the fidelity reduces to
\begin{align}
F\!\left(
\rho_f^{\mathrm{ideal}},
\rho_f^{\mathrm{noisy}}
\right)
&=
\mathrm{Tr}\{
\rho_f^{\mathrm{ideal}}
\rho_f^{\mathrm{noisy}}
\} \nonumber \\
&=
\frac{1}{2^n}
+
e^{\lambda_f}
\sum_i 2^n \alpha_i^2 .
\end{align}
Using \(e^{\lambda_f}\leq 1\), we obtain
\begin{align}
F\!\left(
\rho_f^{\mathrm{ideal}},
\rho_f^{\mathrm{noisy}}
\right)
&\ge
e^{\lambda_f}
\left(
\frac{1}{2^n}
+
\sum_i 2^n \alpha_i^2
\right) \nonumber \\
&=
e^{\lambda_f}.
\end{align}
In the last step, we used the orthogonality of the Pauli basis together with the purity of the ideal state: $\frac{1}{2^n} + \sum_i 2^n \alpha_i^2 = 1$.

Recalling that \(\mathcal V = 1-e^{\lambda_f}\), we arrive at the fidelity lower bound
\begin{equation}
    F\!\left(
    \rho_f^{\mathrm{ideal}},
    \rho_f^{\mathrm{noisy}}
    \right)
    \ge
    1-\mathcal V .
\end{equation}

Beyond noise-informed circuits, the same condition can also be obtained when, for any layer, the decay terms are equal. In this case, the algorithmic outputs are generated by Pauli strings decaying at the same rate, i.e. \(\mathcal{F}_k=\mathcal A_k\). This condition is, for instance, immediately satisfied when the noise is modeled by a uniform depolarizing channel. In the Supplementary Material, we present a more general fidelity bound applicable to analog quantum algorithms under generic Markovian noise models.

\vspace{0.5cm}

\section{Variational quantum algorithms}

\subsection{Noise-induced barren plateaus}

We now demonstrate the application of our formalism to variational quantum algorithms (VQAs)~\cite{Cerezo2021}, focusing specifically on mitigating noise-induced barren plateaus (NIBPs)~\cite{Wang2021}. In a VQA, the quantum circuit is controlled by a set of classical parameters. Following our established notation, each unitary layer $U_k$ is a function of a parameter vector $\vec{\theta}_k$, such that $U_k = U_k(\vec{\theta}_k)$. All trainable parameters across the circuit are then aggregated into a single global vector $\vec{\theta}$. The goal of the algorithm is to optimize these parameters by minimizing a cost function. This function is typically the expectation value of a target Hamiltonian $H$, and is thus expressed as $C(\vec{\theta}) = \Tr{H\rho_f(\vec{\theta})}$, where $\rho_f(\vec{\theta})$ is the final state produced by the parameterized circuit.

The NIBP phenomenon poses a significant challenge to VQAs. It arises when noise in the circuit causes the gradient of the cost function, $\nabla C(\vec{\theta})$, to vanish exponentially with increasing circuit depth~\cite{Larocca2025}, namely the $L$ layers. This occurs because the output state $\rho_f(\vec{\theta})$ is driven towards the maximally mixed state, erasing the landscape features necessary for optimization.

Our analytical framework provides a direct method for mitigating this issue. In particular, while the exponential decay cannot be avoided, as shown in Ref.~\cite{Wang2021}, the presented formalism allows the selection of ansatzes that exhibit a slower decay. Going more into detail, the partial derivative of the cost function with respect to a single parameter $\theta_{k,l}$ (the $l$-th parameter in the $k$-th layer) is given by:
\begin{equation}
    \frac{\partial C}{\partial \theta_{k,l}} = \sum_{i_1, \dots, i_L} \dots \frac{\partial \alpha_{i_{k-1}, i_k}^k}{\partial \theta_{k,l}} \dots e^{(\lambda_{i_1}^1+ \dots + \lambda_{i_L}^L)} \Tr{r_{i_L}^LH}.
\end{equation}
The exponential decay characteristic of NIBPs is captured by the already studied terms $e^{(\sum_j \lambda_{i_j}^j)}$. Consequently, the introduced noise resilience index, $\mathcal{V}$, can also be directly connected to the gradient magnitude. In other terms, the parameterized circuits that minimize $\mathcal{V}$ are also the ones that mitigate NIBPs the most.

\subsection{Hardware-efficient ansatz example}

\begin{figure}[t]
    \centering  
    \includegraphics[width=\linewidth, keepaspectratio]{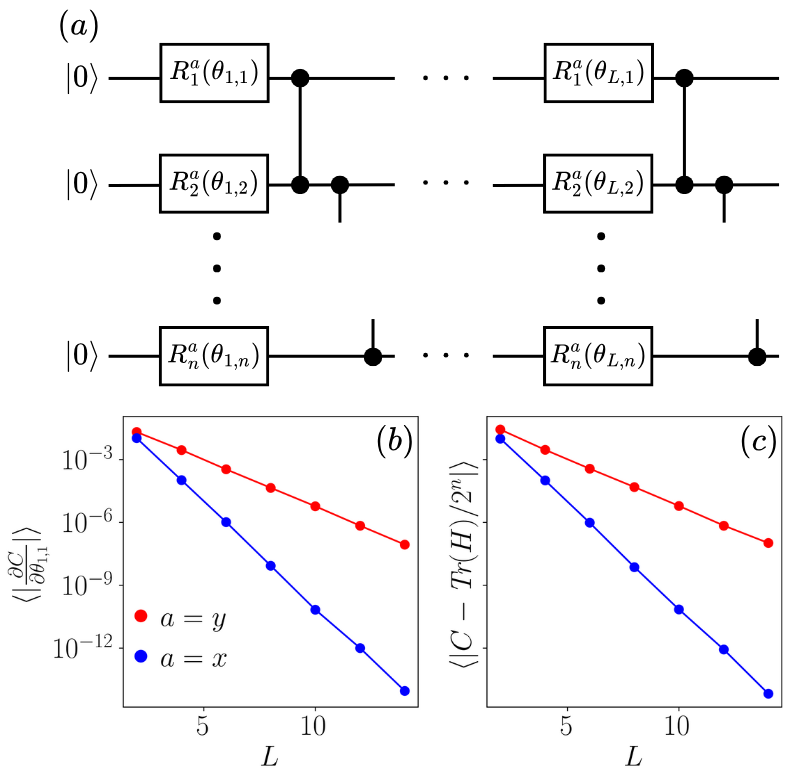}
    \caption{(a) Ansatz used in the numerical simulations. (b) Absolute value of the cost-function derivative with respect to $\theta_{1,1}$. (c) Distance between the cost-function value obtained from the circuit output and the one relative to the fully mixed state. All the points are averages over $10^{4}$ random circuit initializations. Noise parameters are fixed to $q_x = q_z=0.5$, $q_y = 0$, taking $n=8$ and $a$ indicates the orientation of single-qubit rotations gates. A significantly slower decay is observed for the noise-adapted ansatz.}
    \label{fig:Fig_vqa}
\end{figure}

To illustrate our theory, we analyze the optimization of hardware-efficient ansatzes under local noise. As in Ref.~\cite{Wang2021}, we model each circuit layer as the composition of local Pauli channels. After every unitary operation, the noise channel takes the form
\begin{equation}\label{Eq:singlequbit_nm}
    e^{\mathcal{L}_k} = \bigotimes_{j=1}^{n} \mathcal{N}_j,
\end{equation}
where the action of $\mathcal{N}_j$ on the $j$-th qubit is specified by
\begin{eqnarray}
        \mathcal{N}_j(\mathbb{I}) &=& \mathbb{I},\nonumber \\
        \mathcal{N}_j(\sigma_j^a) &=& q_a \sigma_j^a,
\end{eqnarray}
with $\sigma_j^a$ a Pauli operator on site $j$ and parameters $-1 < q_x, q_y, q_z < 1$.

We focus on the anisotropic case where the $y$-direction is maximally affected by noise, i.e., $q_y=0$. Initializing the system with the $n$ qubits in $\ket{\mathbf{0}}$, we study two hardware-efficient variational circuits with building blocks
\begin{equation}\label{Eq:circuit}
    U_k = \prod_{i=1}^{n-1} CZ_{i,i+1} \circ \prod_{i=1}^{n} R^a_{i}(\theta_{k,i}),
\end{equation}
where $CZ_{i,i+1}$ is a controlled-$Z$ gate between qubits $i$ and $i+1$, and $R^a_{i}(\theta_{k,i})$ is a single-qubit rotation of qubit $i$ around axis $a$ [see Fig.~\ref{fig:Fig_vqa} (a) for a representation].

For the chosen initial condition, only the two directions $a=x$ and $a=y$ yield non-trivial outputs. An analysis of the final state Pauli string decomposition for these two cases shows that, for both ansatzes, $\mathcal{V}$ saturates its upper bound: $\mathcal{V} = 1$. However, as discussed in the Supplemental Material, when $a = y$, the number of final right eigenvectors corresponding to this maximum is significantly smaller, making this ansatz noise-adapted according to our noise vulnerability index.

For our numerical analysis, as the cost Hamiltonian, we take $H=\sigma_1^z\sigma_2^z$, and compute both the cost-function derivative and its deviation from the fully mixed state value, $C=\Tr\{H\}/2^n$. As shown in Fig.~\ref{fig:Fig_vqa}, both quantities decay exponentially with the number of layers $L$, confirming the emergence of noise-induced barren plateaus. Importantly, the decay is substantially slower for the noise-aware circuit, demonstrating that such ansatzes allow for an exponentially larger circuit depth at fixed measurement resources.

\section{Experimental noise benchmark in the IBM machine}

\begin{figure}[t]
    \centering  \includegraphics[width=\linewidth, keepaspectratio]{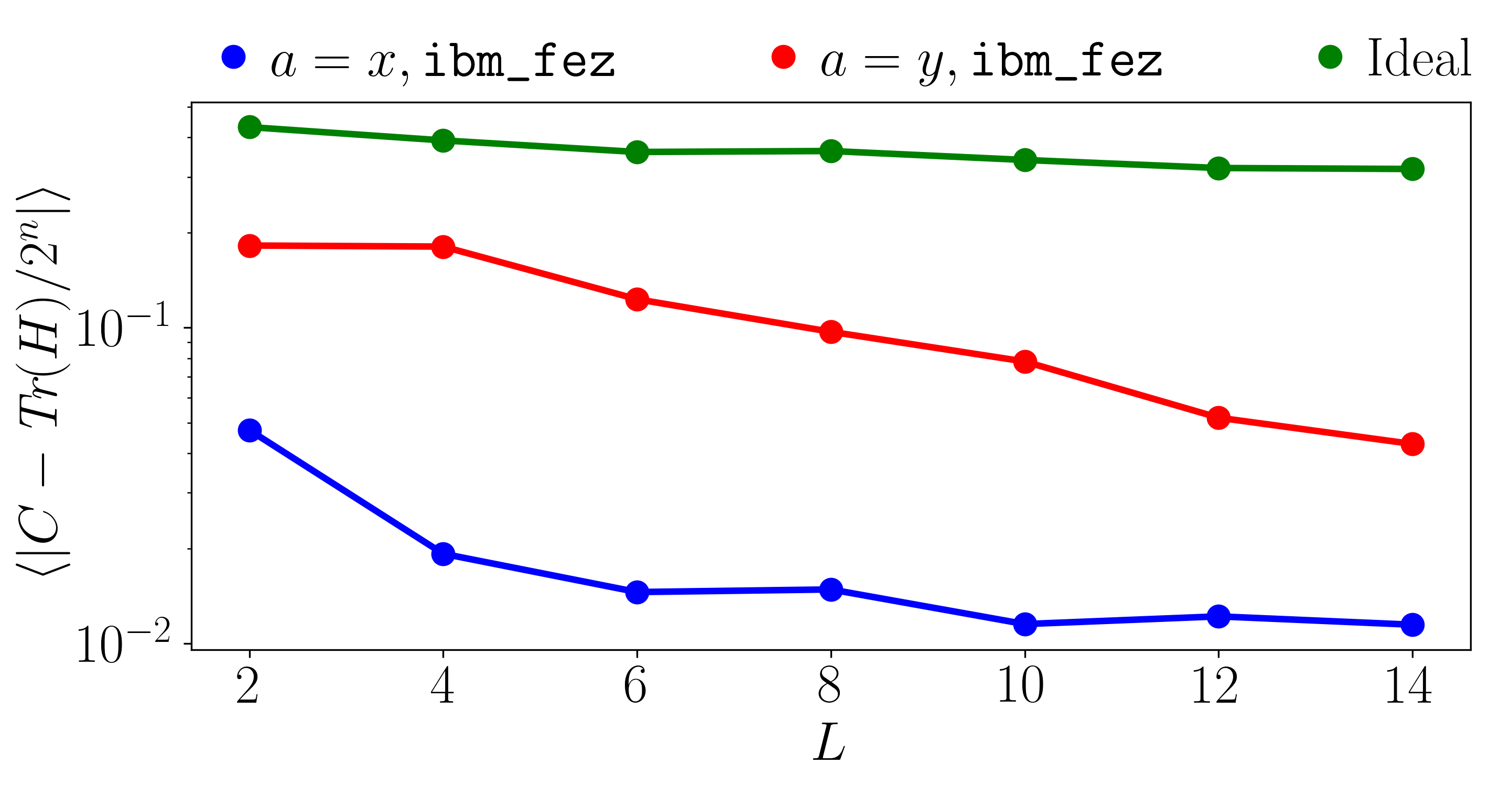}
    \caption{Difference between the observable expectation values evaluated on the circuits implemented on the $\texttt{ibm\_fez}$ device and the one theoretically evaluated on the fully mixed state. Each point is averaged over $100$ random circuit initializations, while each sample is estimated using $10000$ shots. The number of considered device lines used to build both circuits is $n=12$. The green line represents the ideal average distance, which coincides for both circuits while the blue and red lines correspond to different ansatzes associated with distinct orientations, $a$, of the single-qubit rotation gates.}
    \label{fig:IBM}
\end{figure}

Given the framework presented above, the question arises of whether it could be leveraged in a current quantum device. To this end, using the same class of circuits of Eq.~\eqref{Eq:circuit} and the same target Hamiltonian $H=\sigma_1^z \sigma_2^z$, we repeated the previous analysis, implementing the circuits on the 156 qubits $\texttt{ibm\_fez}$ device. For each realization, we estimated the observable expectation value using $10000$ shots. As shown in Fig.~\ref{fig:IBM}, while in the ideal case both circuits take the same average cost function distance from the one evaluated on the fully mixed state, the results on the real device show a clear difference for the two different symmetries. In particular, we observe that the case $a=y$ appears to be more noise resilient, as in the numerical example above. Moreover, we note that the expected exponential decay has not been fully reproduced for $a=x$. We explain this behavior by noting that, given the limited budget in terms of the number of available measurements, the cost function values can be evaluated only up to a precision of the order of $10^{-2}$. Consequently, once the computed difference approaches this limit, we do not have enough experimental resolution to keep reproducing the decay.

We emphasize that, as shown in the Supplemental Material, the information provided in the device documentation~\cite{qiskit}, the characterization of qubit decay times $T_1$ and $T_2$, is insufficient to predict which of the two ansatzes is less vulnerable to noise. Consequently, the experimental results presented here reveal a non-trivial noise feature of the device that cannot be inferred from the standard $T_1$ and $T_2$ characterization alone.

\begin{figure}[t]
    \centering  
    \includegraphics[width=\linewidth, keepaspectratio]{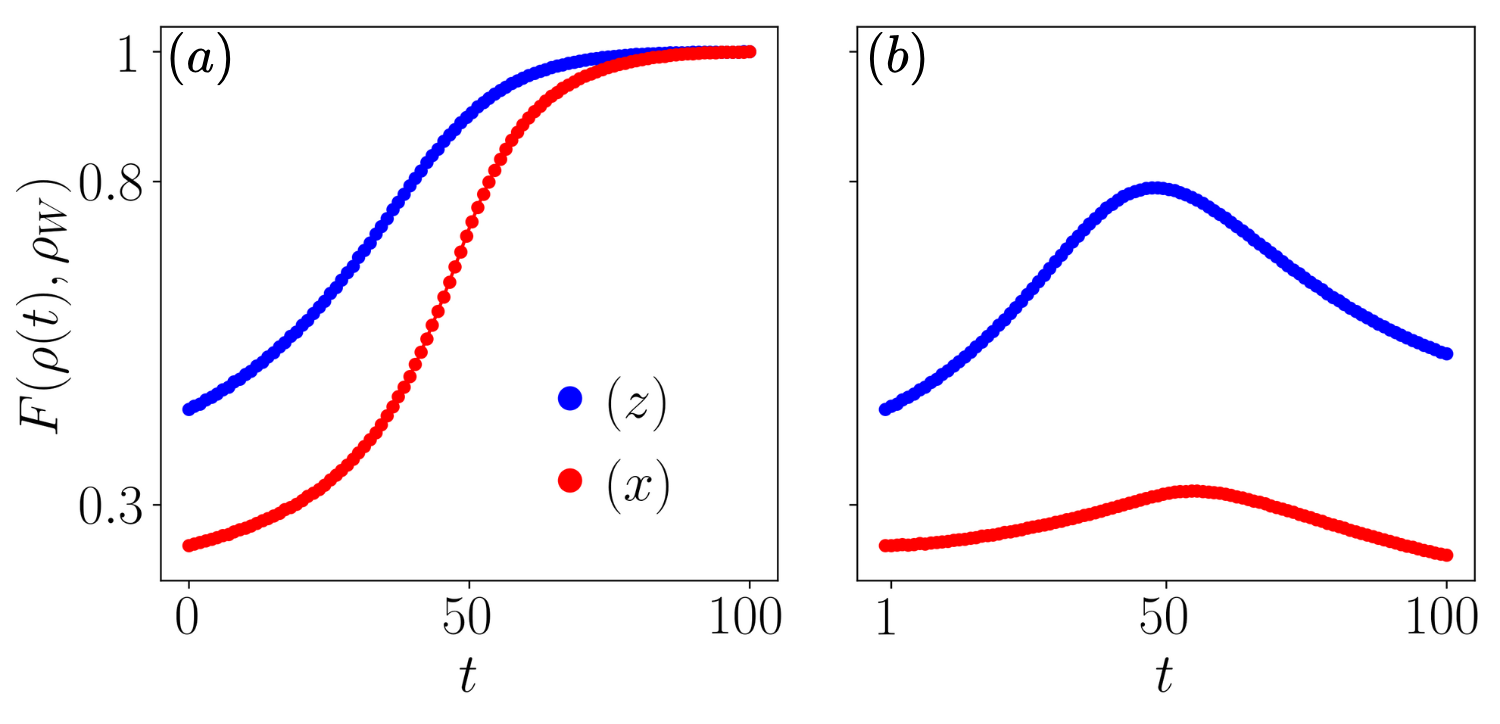}
    \caption{Fidelity evolution over time for the Adiabatic State Preparation example. (a) Noiseless case. (b) Noisy case. 
    The total evolution time is fixed at $T=100$, and the system size is $n=5$. The labels $(z)$ and $(x)$ correspond to the different annealing schedules described in the main text.}
    \label{fig:Fig_ann}
\end{figure}

\section{Adiabatic State Preparation}

As a richer example, we now consider another algorithm belonging to this class: adiabatic state preparation~\cite{asp, Albash2018}. This method relies on the adiabatic theorem, which guarantees that a quantum system initially in the ground state $\ket{\psi_0}$ of a simple Hamiltonian $H_0$ will remain close to the ground state of a changing $H_s$ if such a change is slow enough. By defining $H_f$ as the final Hamiltonian,  whose ground state coincides with the desired target state, the algorithm prepares such a state by evolving $\ket{\psi_0}$ for a sufficiently long time $T$ under the time-dependent Hamiltonian
\begin{equation}
    H(s) = A(s)H_0 + B(s)H_f.
\end{equation}
Here, $s=s(t)$ satisfies $s(0)=0$ and $s(T)=1$. The interpolation functions $A(s)$ and $B(s)$ are chosen so that $A(0)/ B(0) \gg 1$ and $A(1)/ B(1) \ll 1$, ensuring a smooth transition from $H_0$ to $H_f$.

As a concrete instance, we focus on preparing the $n$-qubit W state~\cite{Wstate1, Wstate2, Wstate3}
\begin{equation}
\ket{W}= \frac{1}{\sqrt{n}}\sum_{i=1}^{n} \ket{e_i},
\end{equation}
where $\ket{e_i}$ denotes the computational basis state with all qubits in $\ket{0}$ except the $i$-th one, which is in $\ket{1}$. This state can be obtained adiabatically by taking the target Hamiltonian
\begin{equation}
H_f = -\sum_{i\neq j} \ket{e_i}\bra{e_j},
\end{equation}

whose ground state is, precisely, $\ket{W}$. We study two different initial conditions:
\begin{align}
    (z):\;& H_{0,z} = \sigma_1^z - \sum_{i=2}^n \sigma_i^z, 
    & \ket{\psi_{0,z}} = \ket{e_1}, \nonumber\\
    (x):\;& H_{0,x} = \sigma_1^x - \sum_{i=2}^n \sigma_i^x, 
    & \ket{\psi_{0,x}} = \prod_{i=1}^n R_i^y(\pi/2)\ket{e_1}.
\end{align}

Following the previous methodology, we now examine how these two approaches respond to noise of the form given in Eq.~\eqref{Eq:unb_Lindb}, acting locally on all the qubits. Specifically, we choose
\begin{equation}
\gamma_z = 1/T², \quad \gamma_x = \gamma_y = 0, 
\end{equation}
and we interpolate linearly between $H_0$ and $H_f$ with $A(s)=1-s$ and $B(s)=s$.  

Figure~\ref{fig:Fig_ann}(a) shows the fidelity between the time-evolved state $\rho(t)$ and the target state $\rho_W = \ket{W}\bra{W}$ in the ideal noiseless case. Both initial conditions succeed, achieving final fidelities $F\simeq 1$. However, the situation changes dramatically in the presence of noise, as seen in Fig.~\ref{fig:Fig_ann}(b). For the $(z)$ initialization, fidelities as high as $F\simeq 0.8$ remain achievable, while in the $(x)$ initialization the maximum fidelity drops to $F\simeq 0.3$.  In the Supplementary Material, we demonstrate a direct connection between the introduced noise vulnerability index and the observed fidelity results.

This difference can be understood by noting that, in the $(z)$ case, the Pauli decomposition of $\rho(t)$ predominantly consists of strings commuting with $\sigma^z$, rendering them largely unaffected by the chosen noise channel. In contrast, the $(x)$ case introduces components sensitive to the noise, leading to stronger degradation. Additionally, at long times, a fidelity decay is observed in both cases due to relaxation toward a stationary state imposed by the noise, which generally does not coincide with the target state. Hence, in noisy conditions, one must strike a balance between the total runtime $T$ and the final-state accuracy.

\begin{figure*}[t!]
    \centering  
    \includegraphics[width=\linewidth, keepaspectratio]{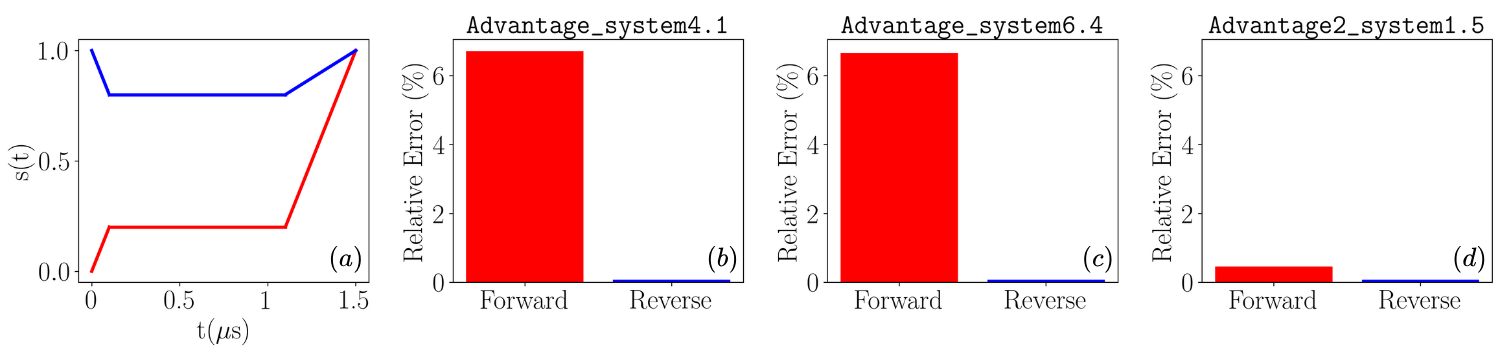}
    \caption{(a) Annealing schedules implemented in the D-Wave devices. 
The forward protocol is shown in red, while the reverse protocol is in blue. 
(b–d) Relative errors between the experimentally measured average energy and the theoretically expected value for the \texttt{Advantage\_system4.1}, 
\texttt{Advantage\_system6.4}, and \texttt{Advantage2\_system1.5} machines, respectively. 
Each experimental point is computed taking $1000$ samples. 
Statistical uncertainties were found to be negligible.}
    \label{fig:Dwave}
\end{figure*}

\section{Experimental noise benchmark in the D-Wave machines}

Analogously to the case of digital algorithms, we now turn to the analog setting and experimentally verify that metastability effects can also be observed in analog quantum processors. 
In particular, we focus on D-Wave's quantum annealing devices~\cite{JohnsonQuantum2011}, which have been used to study both classical as well as quantum spin systems extensively~\cite{KairysSimulating2020, KingQubit2021, KingBeyondclassical2025}.  

We implement quantum annealing experiments on three QPUs spanning two generations of annealers.
For each device, the methodology is as follows: we partition the hardware into two-qubit subsystems, consistent with the native chip connectivity, and apply the same annealing protocol to every pair. This allows us to extract an averaged error profile across the full processor. To probe the role of algorithmic symmetry, we vary the annealing schedule while fixing the target Hamiltonian  
\begin{equation}
    H_f = -0.1(\sigma_1^z + \sigma_2^z) - \sigma_1^z \sigma_2^z.
\end{equation}

We first analyze the standard forward annealing protocol, where the schedule $s(t)$ remains close to zero for most of the evolution, as shown in the red line of Fig.~\ref{fig:Dwave} (a). On D-Wave platforms, every forward anneal starts with 
\begin{align}
    H_0 &= -\left(\sigma_1^x + \sigma_2^x\right), 
    &\quad \ket{\psi_0} = \ket{+}^{\otimes 2}.
\end{align}
Hence, during most of the dynamics, the Pauli decomposition of the state is dominated by strings symmetric under the $\sigma_i^x$ operators.  

As a complementary protocol, we consider reverse annealing, where the schedule starts at $s(0)$=$1$ and $H_0$=$H_f$. 
In this case, we initialize the system in the known ground state of the target Hamiltonian to obtain similar final states for both schedules.
Here, $s(t)$ remains close to one, and the system state predominantly exhibits symmetry with respect to the $\sigma_i^z$ operators, as depicted by the blue line of Fig.~\ref{fig:Dwave}(a).  

To quantify device-induced deviations, we compute the relative error
\begin{equation}
    \text{R.E.}(\%) = 100\times\frac{\bar{E}-E_\text{exp}}{\bar{E}},
\end{equation}
where $E_\text{exp}$ is the experimentally measured average energy and $\bar{E}$ is the theoretically expected energy, obtained from a closed-system classical simulation. 
To compute $\bar E$, we implement classical simulations of closed quantum evolutions corresponding to the protocols considered using the \texttt{QuantumAnnealing.jl} package~\cite{MorrellQuantumAnnealing2024}.
For all cases, we find that $\bar{E} \simeq -1.2$, i.e., the final energy coincides with the ground state energy of $H_f$.
While our implementations are well within the adiabatic regime, we observe small deviations from the ground-state energy arising from the anneal ramps at the beginning and end of the protocols.

The experimental results in Fig.~\ref{fig:Dwave} reveal a systematic asymmetry. For reverse annealing, experimental and theoretical values coincide within numerical precision, yielding vanishing relative errors across all platforms. By contrast, forward annealing shows non-negligible errors, though their magnitude decreases with newer generations of hardware, with the \texttt{Advantage2\_system1.5} device achieving the closest agreement with ideal predictions.  

Taken together, these observations support the interpretation that noise asymmetries in D-Wave processors primarily affect states symmetric with respect to the $\sigma^x_i$ operators. It is worth noting that, in general, the noise models used to describe the behavior of D-Wave machines lead to system thermalization~\cite{Therm1, Therm2, Therm3} in the time regimes considered (of the order of a few hundred microseconds). 
This implies that the decoherence channels related to these platforms are non-unital, in contrast to those considered in the quantum circuit examples. Finally, we emphasize that the noise symmetries identified in our analysis are consistent with previous studies performed in different regimes~\cite{Therm2, Therm3}, thereby supporting the broader generality of this property.

\section{Conclusions}

In this work, we have established a novel approach to witness noise in quantum algorithms by leveraging the phenomenon of metastability in open quantum systems. Our results demonstrate that when quantum hardware noise exhibits metastable dynamics, both digital and analog algorithms can be designed to achieve intrinsic resilience by aligning algorithmic symmetries with the noise structure, without the need for redundant encoding such as in standard quantum error correction.

We provided a general theoretical framework, introduced a practical noise resilience metric that avoids the need for classical simulation of the full algorithm, and illustrated our approach through applications to variational quantum algorithms and analog adiabatic state preparation. Importantly, we validated the relevance of our framework with experimental benchmarks on IBM superconducting processors and D-Wave annealers, confirming that metastable noise effects are present and can be systematically exploited to improve algorithmic outcomes on currently available hardware.

Our findings suggest that the structured properties of noise in quantum devices can serve as a direct algorithmic resource and open new directions for noise-aware algorithm design. This paradigm enables meaningful progress towards robust quantum computation in the NISQ era, bridging theoretical developments and experimental practice. Future work may focus on further characterizing metastability across platforms and extending these concepts to more general noise models that, for example, exhibit features of non-Markovianity.

\vspace{1em}

\section*{Acknowledgements}
The authors acknowledge Davide Venturelli for useful discussions. A. S. acknowledges the Spanish State Research Agency, through the Mar\'ia de Maeztu project CEX2021-001164-M funded by the MICIU/AEI/10.13039/501100011033, through the COQUSY project PID2022-140506NB-C21 and -C22 funded by MICIU/AEI/10.13039/501100011033, MINECO through the QUANTUM SPAIN project, and EU through the RTRP - NextGenerationEU within the framework of the Digital Spain 2025 Agenda. A.S. also acknowledges the CSIC Interdisciplinary Thematic Platform (PTI+) on Quantum Technologies in Spain (QTEP+) and the support of a fellowship from the ``la Caixa” Foundation (ID 100010434 - LCF/BQ/DI23/11990081). A. S. also acknowledges support from the U.S. Department of Energy (DOE) through a quantum computing program sponsored by the Los Alamos National Laboratory (LANL) Information Science $\&$ Technology Institute.
A.S. has also been supported by the USRA Feynman Quantum Academy internship program. 
P.S. acknowledges the support of NNSA for the U.S. DOE at LANL under Contract No. DE-AC52-06NA25396, and Laboratory Directed Research and Development (LDRD) for support through 20240032DR. L.P.G.P. acknowledges support from the Beyond Moore’s Law project of the Advanced Simulation and Computing Program at LANL, and the DOE Office of Advanced Scientific Computing Research, Accelerated Research for Quantum Computing program, Fundamental Algorithmic Research toward Quantum Utility (FAR-Qu) project.
We would also like to thank the New Mexico Consortium, under subcontract C2778, the Quantum Cloud Access Project (QCAP), for providing quantum computing resources and technical collaboration.

\bibliography{bibliography}
\setcounter{section}{0}
\setcounter{equation}{0}
\def\theequation{S\arabic{equation}}
\setcounter{figure}{0}
\onecolumngrid
\section*{Supplemental Material for ``Uncovering and Circumventing Noise in Quantum Algorithms via Metastability''}
\def\thefigure{S\arabic{figure}}

\section{Generalization of the noise vulnerability index to analog algorithms}

In addition to the class of digital algorithms described in the main text, our framework extends naturally to the most general analog setting. In this case, the algorithm is implemented through a time-dependent Liouvillian $\mathcal{L}_I(t)$, and the ideal final state reads  
\begin{equation}
    \rho_f^{\mathrm{ideal}} 
    = \mathcal{T} \exp\!\left( \int_0^T \mathcal{L}_I(t)\, dt \right)[\rho_{\mathrm{in}}],
\end{equation}
where $T$ denotes the total evolution time and $\mathcal{T}$ is the time-ordering operator.

In the presence of noise, $\mathcal{L}_I(t)$ is replaced by the total Liouvillian  
\begin{equation}
    \mathcal{L}_T(t) = \mathcal{L}_I(t) + \mathcal{L}_N(t),
\end{equation}
where $\mathcal{L}_N(t)$ describes an additive noise contribution.

To analyze the noisy evolution, we denote the instantaneous right and left eigenvectors of $\mathcal{L}_N(t)$ by $\{ r_i(t) \}$ and $\{ \ell_i(t) \}$, respectively, and their corresponding eigenvalues by $\{ \lambda_i(t) \}$. Using the Lie–Trotter product formula, the final noisy state can then be written as
\begin{align}\label{Eq:RhoN_cont}
    \rho_f^{\mathrm{noisy}}
    &= \lim_{n \to \infty} 
    \left( \prod_{k=1}^{n}
    e^{\mathcal{L}_N\!\left(t_k\right)\delta t}
    e^{\mathcal{L}_I\!\left(t_k\right)\delta t}
    \right)
    \rho_{\mathrm{in}} \nonumber \\
    &= \lim_{n \to \infty}
    \sum_{i_1,\dots,i_n}
    \alpha_{i_1}^{1}
    \alpha_{i_1,i_2}^{2}
    \cdots
    \alpha_{i_{n-1},i_n}^{n}
    e^{\sum_{k=1}^{n}
        \lambda_{i_k}^k\!\left(t_k\right)
        \delta t}r_{i_{n}}(T) = \lim_{n \to \infty}
    \sum_{i_1,\dots,i_n} \beta_{i_1, \dots, i_n}  e^{\sum_{k=1}^{n}
        \lambda_{i_k}^k\!\left(t_k\right)
        \delta t} r_{i_n}(T),
\end{align}
where $\delta t=T/n$, $t_k=k\delta t$, $e^{\mathcal{L}_I(\delta t)\delta t}[\rho_\mathrm{in}]=\sum_{i_1}\alpha_{i_1}^{(1)}r_{i_1}(\delta t)$, $e^{\mathcal{L}_I(t_k)\delta t}[r_{i_{k-1}}(t_k)]=\sum_{i_k}\alpha_{i_{k-1},i_k}^{(k)}r_{i_k}(t_{k+1})$ and $\beta_{i_1,\dotsc,i_L}=\alpha_{i_1}^1 \alpha_{i_1,i_2}^2 \cdots \alpha_{i_{L-1},i_L}^L$.

This expression generalizes Eq.~\eqref{Eq:RhoN} in the main text and enables an extension of the noise resilience index $\mathcal{R}$ computed in the main text. As already discussed, if the exponentials $\exp\!\left(\sum_{k=1}^n\lambda_{i_k}(t_k)\delta t\right) \to 1$, then $\rho_f^\mathrm{noisy}=\rho_f^\mathrm{ideal}$. Consequently, denoting with $f$ a generic function that maps a time value $t$ to a particular instantaneous eigenvector $\lambda_{i^*}(t)$ appearing in the $\rho_f^\mathrm{noisy}$ expansion, with a corresponding non-vanishing coefficient, we will have the following general expression for the noise vulnerability index:
\begin{equation}\label{Eq:IR_general}
    \mathcal{V} = \mathcal{V}(\mathcal{L}_I) = \max_{f}|1-e^{\int_0^T f(t)dt}|.
\end{equation}

The class of digital algorithms presented in the main text represents a specific instance of this more general analog framework.

\section{Upper bound on the noise vulnerability index in the general setting}

In the most general setting of analog quantum algorithms, we can derive an upper bound on the noise vulnerability index \(\mathcal V\). We assume efficient access, at each time \(t\), to the fastest incoherent decay rates, denoted by \(\gamma_f(t)\). These correspond to the noise eigenvalues $\lambda_i(t)$ with the smallest real parts that affect the algorithmic output. We also assume access to the fastest coherent timescale induced by the noise, denoted by \(\delta_f(t)\), corresponding to the eigenvalue with the largest  imaginary part.

Starting from the definition of \(\mathcal V\) in Eq.~\eqref{Eq:IR_general}, we separate the coherent and incoherent error contributions and write
\begin{equation}
\mathcal V
=
\left|1-e^{\gamma}e^{i\delta}\right|,
\end{equation}
where \(\gamma\in(-\infty,0]\) quantifies the accumulated incoherent decay, while \(\delta\) quantifies the accumulated coherent phase error.

For the sake of clarity, let's assume \(\delta_f(t),\delta \in [0, \pi]\). In this case, \(\mathcal V\) increases monotonically with \(\delta\) and we can bound the total coherent contribution by
\begin{equation}
\delta_M
=
\int_0^T \delta_f(t)\,dt.
\end{equation}

Similarly, the accumulated incoherent part satisfies
\begin{equation}
\gamma
\in
\left[
\int_0^T \gamma_f(t)\,dt,\,
0
\right].
\end{equation}
Once \(\delta_M\) is fixed, upper-bounding \(\mathcal V\) therefore reduces to maximizing
\begin{equation}
\left|1-e^{\gamma+i\delta_M}\right|
\end{equation}
over this interval of admissible values of \(\gamma\). This is a one-dimensional convex optimization problem, and the maximum is attained at one of the interval endpoints. Thus, the optimal incoherent contribution \(\gamma_M\) is given by one of
\begin{equation}
\gamma_M
\in
\left\{
\int_0^T \gamma_f(t)\,dt,\,
0
\right\}.
\end{equation}

We therefore obtain the upper bound
\begin{equation}
\mathcal V
\le
\left|
1-e^{\gamma_M+i\delta_M}
\right|.
\end{equation}
This bound provides a practical criterion for comparing analog quantum algorithms and identifying those that are best suited for implementation on a given noisy device. Finally, the presented calculation can be readily generalized once the interval to which $\gamma$ and $\delta$ belong is known.

\section{General expression for the fidelity bound}
Without any assumption on the quantum algorithm, we now quantify how the noise vulnerability index $\mathcal{V}$ links to the performance of noisy quantum algorithms. Specifically, we will compute the trace distance between $\rho_f^{\mathrm{ideal}}$ and $\rho_f^{\mathrm{noisy}}$, which measures the maximum difference in measurement outcome probabilities between the two states. Firstly, from Eq.~\eqref{Eq:RhoN_cont}, we can expand $\rho_f^{\text{noisy}}$ in a more compact way:

\begin{equation}
\rho_f^{\text{noisy}}=\sum_{i,j}\alpha_{i,j}\, e^{\lambda_j}\, r_i(T),
\end{equation}
where the sum runs over the right eigenvectors $r_i(T)$ and their associated decay terms $e^{\lambda_j}$, and the coefficients $\alpha_{i,j}$ denote the corresponding weights.

Under the assumption of metastable noise, the separation of timescales in the Liouvillian spectrum implies that the slowest decays remain effectively unchanged. This permits approximating the expansion in the following way:
\begin{equation}\label{Eq:Exp_met}
\rho_f^{\text{noisy}}\simeq \sum_i \left(\alpha_{m,i} + \alpha_{d,i}e^{\lambda_m} \right) r_i(T),
\end{equation}
where the $\alpha_{m,i}$ values collect the metastable contributions, the $\alpha_{d,i}$ ones account for the residual coefficient, and $\lambda_m$ is the eigenvalue corresponding to the fastest time-scale in the Eq.~\eqref{Eq:RhoN_cont} expansion.

Crucially, metastability ensures that, in the limit where the faster decaying contributions vanish, i.e. $e^{\lambda_m}\to 0$, the state remains a valid physical one. This property implies the bound
\begin{equation}
|\alpha_{d,i}| \leq 2\big\| \ell_i(T) \big\|_{\infty}.
\end{equation}

The trace distance between the noisy and ideal output now reads:

\begin{equation}
\big\|\rho_f^{\mathrm{ideal}} - \rho_f^{\mathrm{noisy}}\big\|_1
\simeq \Big\| \sum_i \alpha_{d,i}(1-e^{\lambda_m})r_i(T)\Big\|_1 \leq 2\sum_{i} \Big\|\ell_{i}(T) \Big\|_{\infty}\Big\|r_{i}(T)\Big\|_1  \mathcal{V}.
\end{equation}

Moreover, the computed upper bound directly translates to a worst-case fidelity bound:
\begin{equation}
    \sqrt{F(\rho_f^{\mathrm{ideal}}, \rho_f^{\mathrm{noisy}})} \gtrsim 1 - \sum_{i} \Big\|\ell_{i}(T) \Big\|_{\infty}\Big\|r_{i}(T)\Big\|_1  \mathcal{V}.
\end{equation}

We emphasize that the bounds derived above were computed in a manner entirely agnostic to the final outcome of the quantum algorithm. However, incorporating prior knowledge of specific properties of the target state, such as its purity, can be leveraged to obtain a tighter estimate of the fidelity bound, as shown in the main text.

\vspace{0.5cm}

\section{Analytical calculation for the single qubit illustrative example}

We derive the evolution of a qubit undergoing a coherent rotation around the $x$-axis in the presence of the biased Pauli noise described in the main text. The dynamics is governed by the Lindblad master equation
\begin{equation}
\dot{\rho} = -i[H,\rho] + \gamma \left( \sigma^z \rho \sigma^z - \rho \right),
\end{equation}
with Hamiltonian $H = \tfrac{1}{2}\sigma^x$.

Expressing the density matrix in Bloch form,
\begin{equation}
\rho(t) = \frac{1}{2}\left(I + x(t)\sigma^x + y(t)\sigma^y + z(t)\sigma^z\right),
\end{equation}
we obtain the following system of coupled differential equations:
\begin{align}
\dot{x}(t) &= -2\gamma x(t), \nonumber \\
\dot{y}(t) &= z(t) - 2\gamma y(t), \nonumber \\
\dot{z}(t) &= -y(t).
\end{align}

Combining the last two equations yields a closed second-order differential equation for $z(t)$:
\begin{equation}
\ddot{z}(t) + 2\gamma \dot{z}(t) + z(t) = 0,
\end{equation}
which is the equation of a damped harmonic oscillator.

In the underdamped regime ($\gamma < 1$), and for the initial condition $\rho(0)=\ket{1}\bra{1}$, i.e. $z(0)=-1$, $x(0)=y(0)=0$, the solution reads
\begin{equation}
z(t) = - e^{-\gamma t}\left[\cos(\omega t) + \frac{\gamma}{\omega}\sin(\omega t)\right],
\end{equation}
where $\omega = \sqrt{1-\gamma^2}$.

The remaining Bloch components follow as
\begin{equation}
x(t) = 0, \qquad y(t) = -\dot{z}(t).
\end{equation}

In the weak-noise regime ($\gamma \ll 1$), we approximate $\omega \simeq 1$. Evaluating the evolution at $T=\pi$, the final noisy state takes the form
\begin{align}
\rho^N_{f,x}
&\simeq \tfrac12\big(\mathbb{I} + e^{-\gamma T}\sigma^z\big).
\end{align}

\vspace{0.5cm}

\section{Noise vulnerability for the Hardware-efficient ansatzes example}

\subsection{Anisotropic single-qubit Pauli channel}

We now proceed to evaluate the noise vulnerability of the hardware-efficient ansatzes introduced in the main text. Importantly, as required, this calculation does not require knowing the corresponding circuit final states. For the considered noise model, described by Eq.~\eqref{Eq:singlequbit_nm} in the main text, the left and right eigenvectors of the noise superoperator coincide and correspond to the set of Pauli strings. 
Moreover, the noise vulnerability index, $\mathcal{V}$, takes its maximum value whenever the noise channel acts on Pauli strings containing at least one $\sigma^y$ matrix. Consequently, identifying the most noise vulnerable ansatz reduces to counting the number of Pauli strings that contain a $\sigma^y$ operator at each layer of the circuit.

Let us first consider the case $a = y$. The single-qubit rotation gates acting on the initial state $\ket{\mathbf{0}}$ generate superpositions supported on the set of Pauli strings
\begin{equation}
\{\mathbb{I}, \sigma^x, \sigma^z\}^{\otimes n}.
\end{equation}
It is therefore crucial to determine how this set transforms under the entangling layer composed of nearest-neighbor controlled-$Z$ gates,
\begin{equation}
\Bigg(\prod_{i=1}^{n-1} CZ_{i,i+1}\Bigg)
\{\mathbb{I}, \sigma^x, \sigma^z\}^{\otimes n}
\Bigg(\prod_{i=1}^{n-1} CZ_{i,i+1}\Bigg).
\end{equation}

For each pair of neighboring qubits $(i, i+1)$, the controlled-$Z$ gate acts on local Pauli operators according to the following transformation rules:
\[
\begin{array}{c c}
\toprule
\textbf{$P$} & \textbf{$CZ_{i,i+1}\, P \, CZ_{i,i+1}$} \\
\midrule
\mathbb{I}\otimes\mathbb{I} & \mathbb{I}\otimes\mathbb{I} \\
\mathbb{I}\otimes\sigma^x_{i+1} & \sigma^z_i\sigma^x_{i+1} \\
\mathbb{I}\otimes\sigma^y_{i+1} & \sigma^z_i\sigma^y_{i+1} \\
\mathbb{I}\otimes\sigma^z_{i+1} & \mathbb{I}\otimes\sigma^z_{i+1} \\
\addlinespace
\sigma^x_i\otimes\mathbb{I} & \sigma^x_i\otimes\sigma^z_{i+1}\\
\sigma^x_i \sigma^x_{i+1} & \sigma^y_i \sigma^y_{i+1} \\
\sigma^x_i \sigma^y_{i+1} & -\sigma^y_i \sigma^x_{i+1} \\
\sigma^x_i \sigma^z_{i+1} & \sigma^x_i\otimes\mathbb{I} \\
\addlinespace
\sigma^y_i\otimes\mathbb{I} & \sigma^y_i \sigma^z_{i+1} \\
\sigma^y_i \sigma^x_{i+1} & -\sigma^x_i \sigma^y_{i+1} \\
\sigma^y_i \sigma^y_{i+1} & \sigma^x_i \sigma^x_{i+1} \\
\sigma^y_i \sigma^z_{i+1} & \sigma^y_i\otimes\mathbb{I} \\
\addlinespace
\sigma^z_i\otimes\mathbb{I} & \sigma^z_i\otimes\mathbb{I} \\
\sigma^z_i \sigma^x_{i+1} & \mathbb{I}\otimes\sigma^x_{i+1} \\
\sigma^z_i \sigma^y_{i+1} & \mathbb{I}\otimes\sigma^y_{i+1} \\
\sigma^z_i \sigma^z_{i+1} & \sigma^z_i \sigma^z_{i+1} \\
\bottomrule
\end{array}
\]

From this table, we observe that a $\sigma^y$ operator appears on a given qubit whenever a $\sigma^x$ operator in the input string is adjacent to another $\sigma^x$ on a neighboring qubit.  

Therefore, determining whether a Pauli string produces at least one $\sigma^y$ under the full chain of $CZ$ gates is equivalent to checking whether the original string contains two consecutive $\sigma^x$ operators. 
The problem is thus reduced to a purely combinatorial one: counting the number of length-$n$ strings over the alphabet $\{\mathbb{I}, \sigma^x, \sigma^z\}$ that contain at least one pair of adjacent $\sigma^x$ symbols.  

Denoting by $a_n$ the number of strings without adjacent $\sigma^x$ symbols, the number of strings that generate at least one $\sigma^y$ is
\begin{equation}
N_{y}(n) = 3^n - a_n.
\label{Eq:NY}
\end{equation}

The quantity $a_n$ satisfies the recurrence relation
\begin{equation}
a_n = 2 a_{n-1} + 2 a_{n-2}, \qquad 
a_1 = 3, \quad a_2 = 8,
\label{Eq:Recurrence}
\end{equation}
where the first term counts strings beginning with $\mathbb{I}$ or $\sigma^z$, and the second term counts strings starting with $\sigma^x$ followed by a non-$\sigma^x$ symbol.

Finally, since adding further layers does not increase the number of strings associated with the $\mathcal{V}$ maximum, we conclude that the number of right eigenvectors contributing to the evaluation of the noise vulnerability is precisely $N_y$.  

Using analogous arguments, in the complementary case $a = x$, the number of eigenvectors associated with the maximum is equal to all the possible Pauli strings that contain at least one $\sigma^y$: $\tilde{N}_y = 3^n-2^n$. 

For the case studied numerically, $n=8$, we find 
\begin{equation}
N_y = 3217, \qquad \tilde{N}_y = 6305,
\end{equation}
which implies that the ansatz with $a = x$ is more vulnerable to noise as observed.

\vspace{0.5cm}

\subsection{$T_1$/$T_2$ thermal relaxation}

According to the noise model described in the IBM device documentation, the noise acting after each circuit layer can be modeled as a decay process, where the characteristic time values ($T_1$ and $T_2$) are provided in the device specifications~\cite{qiskit}. We now demonstrate that this information alone is insufficient to determine which of the two proposed hardware-efficient ansatzes is more resilient to noise. In particular, our experimental results reveal noise features that are not directly accessible from publicly available device data.

We denote with $t_g$ the time of each unitary operation, assumed, for the sake of clarity, to be constant across all layers. Within this model, each Liouvillian $\mathcal{L}_k$ is written as a sum of local contributions,
\begin{equation}
\mathcal{L}_k = \sum_i \mathcal{L}_i t_g,
\end{equation}
where the  $\mathcal{L}_i$ acts on the $i$-th qubit and is given by
\begin{equation}
\mathcal L_i[\rho]
=
\frac{1}{T_1}
\left(
\sigma^- \rho \sigma^+
-
\frac12 \{\sigma^+\sigma^-,\rho\}
\right)
+
\frac{1}{2T_\phi}
\left(
\sigma^z \rho \sigma^z - \rho
\right),
\end{equation}
with the relation
\begin{equation}
\frac{1}{T_2} = \frac{1}{2T_1} + \frac{1}{T_\phi}.
\end{equation}

In the operator basis $\{\mathbb{I},\sigma^x,\sigma^y,\sigma^z\}$, the operators $\sigma^x$ and $\sigma^y$ are eigenoperators of the Liouvillian, both with eigenvalue
\begin{equation}
\lambda_x = \lambda_y = -\frac{1}{T_2}.
\end{equation}

The remaining orthogonal subspace is spanned by right eigenvectors whose decay rates are governed by $1/T_1$.

Consider a Pauli string $\mathcal{P}$ containing $n_x$ factors of $\sigma^x$, $n_y$ factors of $\sigma^y$, and $n_{Iz}$ factors in the $\{\mathbb{I},\sigma^z\}$ sector (identities or $\sigma^z$ operators), with $n_x + n_y + n_{Iz} = n$, decomposing $\mathcal{P}$ into right eigenvectors of the Liouvillian $\mathcal{L}_k$, the maximal decay rate is
\begin{equation}
\lambda_{\text{max}}(\mathcal{P})
=
-\frac{(n_x + n_y)t_g}{T_2}
-\frac{n_{Iz}t_g}{T_1}.
\end{equation}

while the minimum is 
\begin{equation}
\lambda_{\text{min}}(\mathcal{P})
=
-\frac{(n_x + n_y)t_g}{T_2}.
\end{equation}

The two hardware-efficient ansatzes differ only by replacing every $\sigma^x$ with $\sigma^y$ in the Pauli-string decomposition at each layer. This substitution leaves both $n_x + n_y$ and $n_{Iz}$ unchanged for every Pauli string. Consequently, the two ansatzes experience identical decay under this noise model.

For simplicity, assuming that the $T_1$ and $T_2$ times are uniform across all qubits, the equivalence between the two ansatzes is also reflected in the noise vulnerability index:
\begin{equation}
\mathcal{V}_y = \mathcal{V}_x
= 1 - \exp(-nt_g/T_2),
\end{equation}
showing that, within a pure-decay model, both ansatzes exhibit the same level of noise vulnerability.

This conclusion, however, does not rely on the assumption of uniform decay rates. In the more general case where $T_1^{(i)}$ and $T_2^{(i)}$ vary from qubit to qubit, the noise vulnerability index is obtained by replacing $nt_g/T_2$ with the corresponding weighted sum of local decay rates. Since the substitution $\sigma^x$ with $\sigma^y$ preserves the number of operators in the ($x,y$) sector on each qubit, the two ansatzes remain equivalent under any local decay channel.

Consequently, the experimental data presented in the main text reveal a noise feature that is not captured by a standard lifetime characterization alone.

\section{Noise vulnerability for the adiabatic state preparation example}

Similarly to what was done for the quantum circuit examples, we analyze the adiabatic state preparation algorithm by exploiting the fact that the noise channel is diagonal in the Pauli basis. Because of this property, studying the Pauli decomposition of the algorithm outputs over time allows us to estimate the noise vulnerability of the two algorithms under consideration. As before, analyzing how the Hamiltonian dynamics modifies the support of the Pauli strings provides the answer.

Starting with the $(z)$ example, we observe that the initial qubit state is generated by Pauli strings belonging to the set $\{\mathbb{I}, \sigma^z\}^{\otimes n}$. After the action of the Hamiltonian
\begin{equation}
H(s) = A(s)H_0 + B(s)H_f ,
\end{equation}
we note that only the action of $H_f$ contributes to the generation of new Pauli strings in the state decomposition over time. In particular,
\begin{equation}
H_f\{\ket{e_i}\}_i = \{\ket{e_i}\}_i ,
\end{equation}
which implies that the only additional Pauli strings that can appear in the decomposition are those composed of $\mathbb{I}$ and $\sigma^z$, together with at most two operators among $\sigma^x$ and $\sigma^y$.

As a consequence, for an execution time $t$ of the algorithm, the vulnerability index takes the form
\begin{equation}
\mathcal{V}_z(t) = 1-\exp\!\left(\int_0^t -\frac{4}{T}\,dt \right)
       = 1-\exp\!\left(-\frac{4t}{T}\right).
\end{equation}

For the $(x)$ algorithm, the situation is different. The initial state already contains the fastest-decaying terms, corresponding to Pauli strings of the form $\bigotimes_{i=1}^n \sigma_i^x$. Since the exact coefficients of the Pauli decomposition depend on the full time evolution of the state, and are not known without explicitly solving the dynamics, we cannot exclude the presence of these terms at all times during the algorithm. Consequently, we obtain the following worst-case bound for the noise vulnerability:
\begin{equation}
\mathcal{R}_x(t) \leq 1-\exp\!\left(-\frac{2nt}{T}\right).
\end{equation}

Following the approach used in the main text, we quantify the algorithm performance through the fidelity with respect to the target final state $\rho_W = \ket{W}\bra{W}$:
\begin{equation}
F(\rho(t), \rho_W) = \frac{1}{n}\sum_{i,j}\bra{e_i}\rho(t)\ket{e_j}.
\end{equation}

We observe that the diagonal elements in this sum are unaffected by the noise:
\begin{equation}
\mathcal{L}^{\dagger}(t)[\ket{e_i}\bra{e_i}]
= i[H(t), \ket{e_i}\bra{e_i}]
+ \gamma_z \mathcal{D}_z[\ket{e_i}\bra{e_i}]
= i[H(t), \ket{e_i}\bra{e_i}],
\end{equation}
which corresponds exactly to the noiseless evolution.

As a consequence, the performance degradation of the algorithm can be expressed as
\begin{equation}
F(\rho^{\text{ideal}}(t), \rho_W)
-
F(\rho^{\text{noisy}}(t), \rho_W)
=
\frac{1}{n}\sum_{\substack{i,j\\ i\neq j}}
\bra{e_i}
\big(\rho^{\text{ideal}}(t)-\rho^{\text{noisy}}(t)\big)
\ket{e_j}.
\end{equation}

According to Eq.~\eqref{Eq:Exp_met}, for studying the initial noise effects, this difference can be bounded as
\begin{equation}
F(\rho^{\text{ideal}}(t), \rho_W)
-
F(\rho^{\text{noisy}}(t), \rho_W)
\lesssim
\frac{2}{n}\sum_{\substack{i,j\\ i\neq j}}
|\bra{e_i}\rho^{\text{ideal}}(t)\ket{e_j}|\, \mathcal{V}(t),
\end{equation}
noticing that that the vectors $\ket{e_i} \bra{e_j}$ are eigenvectors of the noise model.

From this expression, observing the exponential decay behavior of $\mathcal{V}(t)$ for both algorithms, we can estimate a characteristic time scale $t^*$ beyond which performance degradation is expected to appear.

In the two cases, substituting the values used in the numerical examples, $n=5$ and $T=100$, we obtain
\begin{align}
    t_z^* &= \frac{T}{4} = 25, \nonumber\\
    t_x^* &= \frac{T}{2n} = 10,
\end{align}
which is consistent with the behavior observed in the numerical results presented in the main text.

\end{document}